# The Very Red Halo of the cD Galaxy in A3284 [*]

**E. Molinari[1], A. Buzzoni[1], G. Chincarini[1,2], and M.D. Pedrana[1]**

[1] Osservatorio Astronomico di Brera, Via Brera, 28  20121 Milano, Italy
[2] Università degli Studi, Via Celoria, 16  20133 Milano, Italy
    e-mail internet: molinari at astmim.mi.astro.it

**Abstract.** We present deep CCD surface photometry in the Gunn $g, r, i$ system and spectroscopy of the cD galaxy at the centre of the cluster A3284 at $z = 0.15$.

The brightness profile of the galaxy has been tracked up to 40 arcsec (142 kpc) from the centre, and down to 26 mag/arcsec². The core and halo components in the galaxy have been singled out deriving geometrical parameters of the fitting isophotes as well as magnitudes and colours.

The spectral properties of the galaxy core indicate a stellar population super metal-rich with [Fe/H] = +0.5.

The cD halo is clearly dominant at 45 kpc from the galactic centre and has exceedingly red colours ($g - r = 1.03$, $g - i = 1.82$), about 0.7 mag redder than the core $g - i$. A match with the models for evolutionary population synthesis by Buzzoni (1989) show that the halo is consistent with a population of unevolved M-dwarf stars lower than 0.7 $M_\odot$. The M/L ratio in $B$ for the halo is estimated to range between 50 and 200 implying a total mass for the cD galaxy of $1.6$-$3.1$ $10^{13} M_\odot$ and a total $B$ luminosity of $6.0$ $10^{11} L_\odot$.

**Key words:** galaxies: cd, evolution, photometry, stellar content

## 1. Introduction

A complete study of the cluster evolution must deal with the relevant phenomenon of cD galaxies. As first recognized by Matthews *et al.* (1964), these objects are often present in the centre of rich clusters and display a shallower but very extended envelope that embodies the core and clearly breaks from the galaxy radial luminosity profile up to distances of about 1.5 Mpc (Shombert 1988).

Current theories trying to explain the origin of cD galaxies basically branch in three distinct families. Merrit (1984) argued that the giant envelope might simply be a relic of the original proto-galaxy and therefore it should be the prevailing structure in the process of galaxy formation. Contrary to the other galaxies, the central location of the proto-cD should have minimized any disruptive action generated by the cluster gravitational potential, preserving the external envelope up to the present epoch.

On the other hand, the origin of the halo might reside in accretion mechanisms able to strip matter from the other nearby galaxies via tidal interactions and dynamical friction (Malumuth and Richstone 1984) and/or accrete gas onto the proto-cD via radiative cooling flows of the uncollapsed intracluster medium (Fabian *et al.* 1982).

Mergers can also be thought to be progenitors of the giant ellipticals as N-body simulations lead to predict (Barnes 1989; Governato *et al.* 1991).

In the course of our observing programme on distant clusters of galaxies (Molinari *et al.* 1990, 1993) we obtained multicolor surface photometry of the giant cD galaxy in the center of A3284 at $z = 0.15$. A detailed study of the cluster galaxy population is presented in Molinari *et al.* (1993) to which we refer the reader for any further detail in the data acquisition and reduction.

Here we present the analysis of the surface photometry and the spectra obtained for the cD.

## 2. Observations

The central field of A3284 was imaged in the Gunn $g, r$, and $i$ system during an observing run using the EFOSC at the cassegrain focus of the ESO 3.6m telescope et La Silla (Chile) in November 1986. We used the 512×320 RCA CCD # 8 giving a scale of 0.675 arcsec/pixel and a field of 5.7×3.6 arcmin.

In a second run in November 1987 we used EFOSC for multi-object spectroscopy gathering spectra for eight galaxies members of the cluster in addition to the cD. Figure 1 shows a picture of the cD field in the Gunn $r$ filter while Table 1 gives a summary of the log of observations.

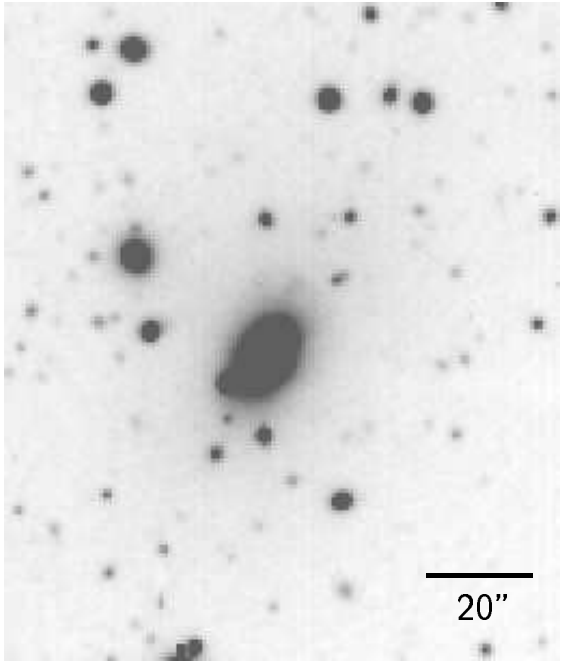

**Fig. 1.** A 20 min $r$ picture of the cD galaxy in A3284 obtained at the ESO 3.6m telescope equipped with EFOSC. North is up, East to the right.

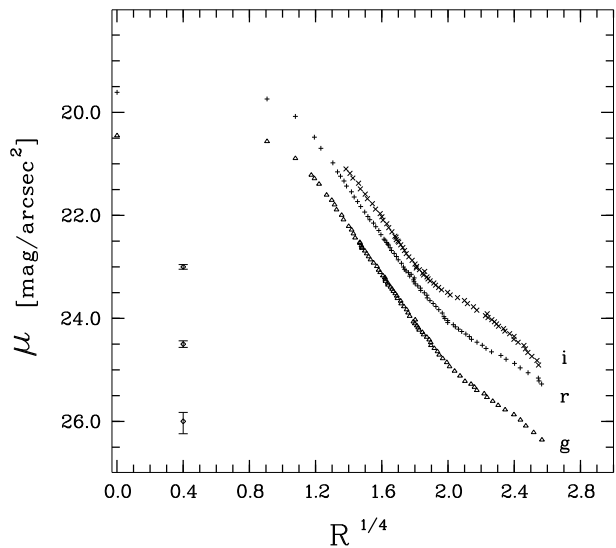

**Fig. 2.** cD surface-brightness profiles in the Gunn $g, r,$ and $i$ bands. Distance from the centre is defined as $R = \sqrt{a \times b}$ in arcsec [1 arcsec = 3.5 kpc assuming $(H_o, q_o) = (50$ km/sec/Mpc, 0)]. Displayed are also the error bars including the uncertainty in the sky subtraction, photon noise and photometry zero point. The galactic halo becomes evident beyond 13 arcsec from the nucleus ($R^{1/4} \geq 1.9$).

### 2.1. Surface Photometry

Image processing was accomplished according to the procedure described in Molinari et al. (1990, 1993). This led

**Table 1.** Observation log

| filter | date | $t_{exp}$ [sec] | seeing [arcsec] | sky rms |
|--------|------|-----------------|------------------|---------|
| g | Nov 4 1986 | 1200 | 2.1 | 0.9% |
| r | Nov 4 1986 | 900 | | |
| r | Nov 4 1986 | 300 | | |
| r | coadded | image | 2.0 | 0.6% |
| i | Nov 4 1986 | 1800 | 1.9 | 0.9% |
| MOS | Nov 21 1987 | 1800 | | |
| MOS | Nov 21 1987 | 2400 | | |

to a final sky rms measured on single pixels, always less than 1%.

The core of the cD within 3 arcsec was saturated in the $i$ image. This, however, has not caused a serious lack of information in the photometry since also in the other bands the core profile was strongly affected by the poor seeing conditions (cf. Table 1).

For the calibration we used eight Feige and Landolt standard stars by translating Johnson magnitudes to the Gunn system via the equations of transformation in Molinari et al. (1990). A conservative uncertainty estimate of magnitude zero-points is ±0.08 mag for all the three $g, r$ and $i$ filters.

To measure the cD surface brightness we used the algorithm SURFPHOT devised by Valentijn and implemented in the package MIDAS. The method consists of fitting elliptical isophotes at fixed intensity levels on the galaxy image and recording then the geometrical parameters. As it is well known, the exact determination of the sky level was found to be the major source of uncertainty in the profile evaluation (Shombert 1988; Franx et al. 1989). We measure a value of 0.5% of the average sky level for the deviation from flatness after the flat-fielding operation. Since we have no way to determine the true background shape we consider this deviation as our main source of error, being the sky rms much smaller when averaged on the large area of the cD halo (rms is ≤ 1% on single pixels).

The final $g, r,$ and $i$ surface-brightness profiles of the cD are reported in Fig. 2 as a function of the radial distance $R^{1/4}$ intended as the geometric mean of the major and minor axis of the isophotes ($R = \sqrt{a \times b}$) in arcsec. Apart from the seeing-induced flattening in the first 2 arcsec from the centre, the de Vaucouleurs $R^{1/4}$ profile is evident until the onset of the halo, which appears at a distance of $R = 13$ arcsec corresponding to 45 kpc by assuming $(H_o, q_o) = (50$km/sec/Mpc, 0). The 1-$\sigma$ error bars in the photometry as a function of the surface brightness are also plotted in Fig. 2. They account both for the scatter in the sky level, the Poissonian photon noise in the signal and the estimated error of 0.08 mag in the absolute photometry zero point.

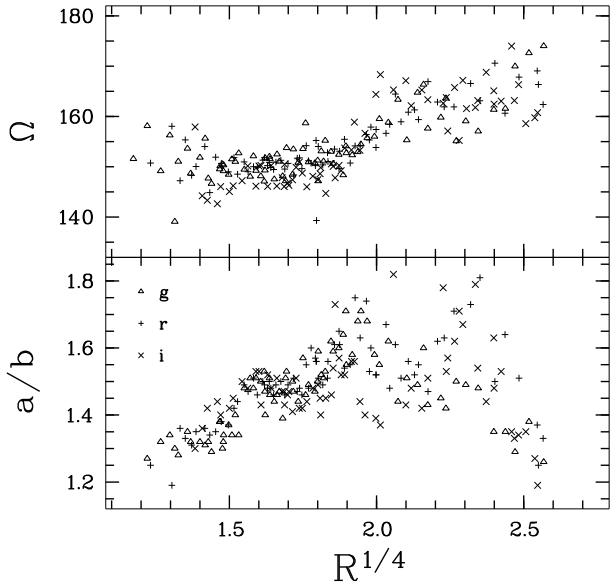

**Fig. 3.** Position angle and axis ratio for the fitting elliptical isophotes as a function of the distance from the cD center. Note the change in the trend at the onset of the halo, about $R^{1/4} \sim 1.9$.

In Fig. 3 we show the geometrical isophotal parameters, plotting the radial trend of the position angle $\Omega$ of the fitting ellipses and the flattening ratio $a/b$. It is remarkable to note that at the halo onset (see Fig. 2 at about $R^{1/4} \sim 1.9$) the position angle abruptly turns by 15° while isophote ellipticity becomes roughly constant. This might be a sign of a peculiar dynamical status of the core/halo system.

By subtracting from the brightness profiles of Fig. 2 the $R^{1/4}$ component (this was accomplished by means of a linear fit in the range $1.5 \leq R^{1/4} \leq 1.8$) we were able to single out the contribution from the halo and derived total magnitudes and colours. These are shown in Table 2 together with other relevant photometric parameters. In the table, magnitudes and colours are integrated up to 40 arcsec from the centre (i.e. 142 kpc with $H_0 = 50$ km/sec/Mpc and $q_0 = 0$) and the errors quoted therein include all relevant contributions: zero point in the photometry, uncertainty in sky subtraction and poissonian noise. Due to the problems with the central saturation, the total $i$ magnitude is less reliable and has been derived from $g$ and $r$ by assuming the mean colours of the range $1.5 \leq R^{1/4} \leq 1.8$ and no colour gradient in the core (cf. Fig. 2).

Effective radius $r_e$ and surface brightness $\mu_e$ reported in the table come from the de Vaucouleurs fit ($\mu = \mu_e + 1.332[(r/r_e)^{1/4} - 1]$) and refer therefore to the underlying elliptical alone. The cD effective radius and surface brightness (i.e. for the whole system elliptical+halo) are quoted as $r'_e$ and $\mu'_e$ and are computed by considering the total radial profile.

### 2.2. Spectroscopy

In addition to eight member galaxies of A3284, the central region of the cD was also observed spectroscopically during the Nov 1987 run with EFOSC using the ESO grism B300 with a dispersion of 230 Å/mm. Two frames were taken for a total of 70 min exposure time by centering E/W a 24×2 arcsec slit on the cD nucleus and with a resolution of 14.4 Å FWHM. Wavelength calibration was performed through He/Ar-lamp exposures while the standard star EG54 served for the relative flux calibration.

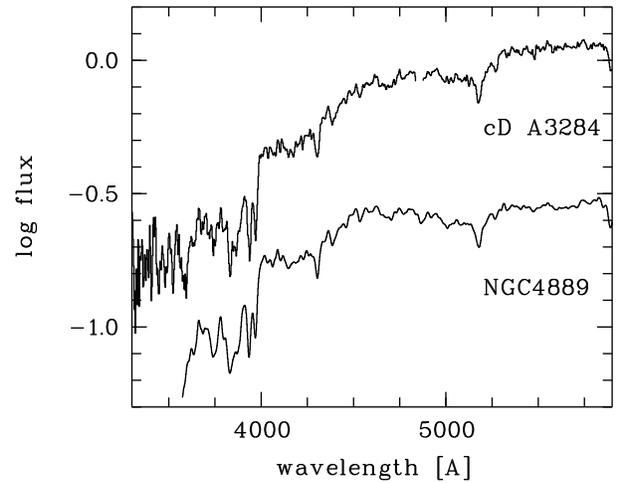

**Fig. 4.** Coadded restframe spectrum of the cD compared with that of N4889, a giant elliptical galaxy in the Coma cluster (from Kennicutt 1992, degraded to our resolution of 14.4 Å/(1+z) = 12.5 Å).

Recognition of the more prominent absorption features in the spectra allowed to derive the redshift of the cD that resulted in $z = 0.149 \pm 0.001$ in good agreement with the mean redshift of the cluster, $z = 0.150 \pm 0.001$ (Molinari *et al.* 1993).

Unfortunately, the exceedingly low surface brightness forced us to integrate the spectrum over the whole spatial coordinate preventing any study of possible radial trends in the spectral features. Our $S/N$ is $\sim 50$ in the galaxy core while at a distance of 10 arcsec from the nucleus the $S/N$ quickly degrades to $\sim 5$ and did not allow us any reliable observation of the halo.

The coadded spectrum of the cD galaxy, corrected for redshift and obtained by stacking the two partial frames and by integrating over the spatial coordinate, is shown in Fig. 4. For comparison, also the spectrum of N4889 a giant elliptical in the Coma cluster (from Kennicutt 1992) is displayed in the figure.

It is remarkable to note a substantially similar spectral distribution of the two galaxies. The A3284 cD is only slightly redder while features like those of MgH at 5180 Å, the G-band at 4300 Å, and the Calcium H+K lines at 3950 Å appear with the same strength in the two spectra. For N4889, Faber *et al.* (1989) report $B - V = 0.99$ and $Mg_2 = 0.359$. On the basis of the $Mg_2$ calibration by Buzzoni *et al.* (1992) this leads to attribute to this galaxy a metallicity [Fe/H] $\sim +0.5$ (i.e. about three times the solar value), and this value is to be regarded also as a fair estimate for the metallicity of the A3284 cD galaxy.

**Table 2.** Photometric parameters

| | $r_e$ [arcsec] | $\mu_e$ [mag arcsec$^{-2}$] | $r'_e$ [arcsec] | $\mu'_e$ [mag arcsec$^{-2}$] | $m_T$(core) | $m_T$(halo) | $m_T$(cD) |
|---|---|---|---|---|---|---|---|
| g | 10.3($\pm$0.4) | 24.04($\pm$0.10) | 11.1($\pm$0.6) | 24.20($\pm$0.04) | 15.83($^{-0.10}_{+0.11}$) | 17.39($^{-0.22}_{+0.31}$) | 15.60($^{-0.09}_{+0.09}$) |
| r | 11.7($\pm$0.7) | 23.52($\pm$0.16) | 12.9($\pm$1.0) | 23.67($\pm$0.06) | 15.01($^{-0.11}_{+0.11}$) | 16.36($^{-0.24}_{+0.29}$) | 14.73($^{-0.10}_{+0.10}$) |
| i | 11.0($\pm$0.9) | 23.07($\pm$0.24) | 15.1($\pm$1.5) | 23.40($\pm$0.09) | 14.70($^{-0.14}_{+0.15}$) | 15.57($^{-0.25}_{+0.32}$) | 14.30($^{-0.11}_{+0.12}$) |
| g–r | | | | | 0.82($\pm$0.15) | 1.03($\pm$0.37) | 0.87($\pm$0.13) |
| g–i | | | | | 1.13($\pm$0.18) | 1.82($\pm$0.39) | 1.30($\pm$0.15) |

## 3. Discussion

A $g - r$ vs. $g - i$ diagram for the 203 objects in the field of A3284 with complete photometry from Molinari *et al.* (1993) is presented in Fig. 5. Looking at the diagram, a main feature is the concentration of points with average colours $(g - r, g - i) = (0.68, 1.04)$. As discussed also in Buzzoni *et al.* (1993) this clump marks the *bona-fide* elliptical galaxy population in the core of A3284.

Located to slightly redder colors in the diagram $[(g - r, g - i) = (0.82, 1.13)]$ with respect to the clump of the elliptical galaxies is the cD core. This is in agreement with its expected larger metallicity, and as shown in Molinari *et al.* (1993; their Fig. 10a) the cD also obeys the canonical Visvanathan-Sandage (1977) c-m relation observed for the early-type galaxy population of A3284.

The halo colours, integrated from 13 to 40 arcsec from the nucleus, are also displayed in Fig. 5 with its 1-$\sigma$ confidence level. It is immediate to see that the halo is much redder than the mean colours of the cluster ellipticals as well as than those of the cD core to a 2-$\sigma$ confidence. Indeed, we are inclined to believe that the redder nature of the halo component of the cD galaxy in A3284 is real, and resembles to some extent the unique case of the giant elliptical in the X-ray selected cluster 1E1111.9-3754 found by Maccagni *et al.* (1988).

The intrinsic colours inferred for the cD halo closely match those of stars of late spectral type about K8-M0, and we can firmly rule out the occurrence of a so large internal reddening in the galaxy. A possibly viable solution to explain such peculiar photometric properties would require a stellar population strongly dominated by low-mass stars. The cooling flows scenario, as stated by Fabian *et al.* (1982) suggest a lowering of the Jeans mass in physical conditions that prevent the formation of dust grains, although the mass of an M0 star ($\sim 0.6 M_\odot$) may be too low to be considered as a typical mass for the star formation in accreted gas.

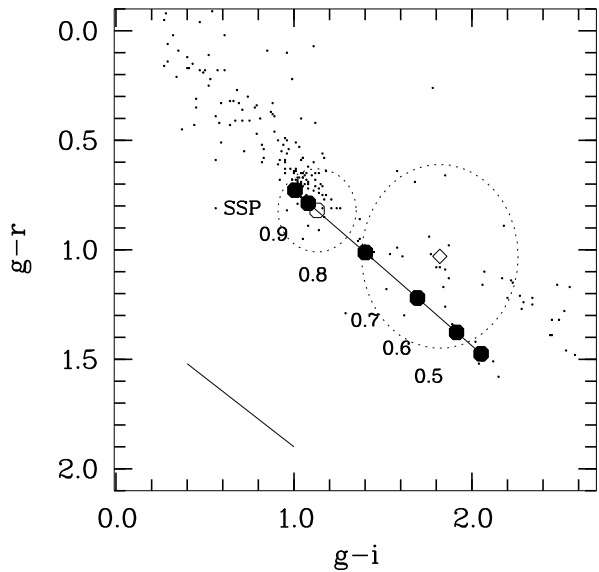

**Fig. 5.** The cD location in the $g - r$ vs. $g - i$ color plane. The open dot marks the colours of the galaxy after removing the halo contribution while the losange shows the integrated colours for the halo. The dashed regions are the 1-$\sigma$ confidence level. Small dots display the 203 objects in the field of A3284 with complete photometry from Molinari *et al.* (1993). Note the clump of cluster elliptical galaxies about $(g - r, g - i) = (0.68, 1.04)$. Bottom left is the reddening vector for E(B-V) = 0.3 mag. The solid curve is the locus for a 15 Gyr Salpeter SSP with [Fe/H] = +0.22 by truncating the upper stellar mass limit from 0.9 to 0.5 $M_\odot$ in steps of 0.1 $M_\odot$ as labelled in the sense of increasing $g - i$. The bluest colours along the curve mark the complete SSP (that is by fully accounting for its Post-Main Sequence evolution).

The possibility of dwarf-dominated stellar populations as a result of galactic cooling flows has been explored by Silk *et al.* (1986) by means of Bruzual's (1983) models for population synthesis. The main conclusion of that work was that there were no conclusive evidences leading to prefer low-mass stars as main contributors to the dark haloes around elliptical galaxies.

A test in this regard was performed by using the Buzzoni (1989) models for a 15 Gyr Salpeter simple stellar population (SSP) of super-solar metallicity ([Fe/H] = +0.22). In the models, the upper cut off of the Salpeter IMF was truncated from 0.9 $M_\odot$ down to 0.5 $M_\odot$ in steps of 0.1 $M_\odot$ while the lower stellar mass limit was always set to 0.1 $M_\odot$. The results are shown in Fig. 5, together with the case of the complete SSP (with a Turn Off stellar mass of 1.00 $M_\odot$). It can be seen that the observed colours of the halo cannot be explained by reddening and support, or at least are not in disagreement with, a stellar population that would basically consist of unevolved dwarfs with $M \leq 0.7 M_\odot$.

This value can be regarded as an upper limit as in no way models using complete SSPs are able to match the observations even by acting drastically on distinctive parameters like the IMF slope, metallicity or age.

Assuming that our SSP truncated at 0.7 $M_\odot$ is a suitable fit to the data we derive $M/L_B = 200$ (in solar unit) for the stellar population in the cD halo. This value strongly depends however on the lower stellar mass limit assumed. In fact, with a Salpeter IMF (and even more for a steeper slope) it happens that the upper Main Sequence mainly contributes to the total light while the lower Main Sequence mostly contributes to the total mass of the population. By simply assuming a M/L ratio pertinent to M0 dwarfs, one would derive $M/L_B \sim 50$ (e.g. Allen 1973).

The estimated $B - g$ colour in the restframe for the halo fitting SSP is found to be $B - g = 1.00$ while the k-correction to $g$ magnitudes accounting for the cD redshift is $k(g) = 0.63$ mag. For the underlying elliptical, a 15 Gyr Salpeter SSP with [Fe/H] = +0.22 could be a suitable fit. In this case we have $B - g = 0.71$, $k(g) = 0.37$ mag and $M/L_B = 21.9$ obtained by integrating the IMF up to 100 $M_\odot$ to account also for dead stars. For A3284 we derived a distance modulus $M - m = 39.9$ assuming $(H_o, q_o) = (50, 0)$ (Molinari *et al.* 1993).

With these numbers, and from the relevant data in Table 2, we can finally compute the absolute $B$ magnitude of the cD core and halo in the restframe:

$$M_B = 15.8 - 39.9 - 0.4 + 0.7 = -23.8 \quad \text{(core)},$$

$$M_B = 17.4 - 39.9 - 0.6 + 1.0 = -22.1 \quad \text{(halo)},$$

from which we derive a $B$-luminosity $L_B(core) = 5.0 \ 10^{11}$ $L_\odot$ and $L_B(halo) = 1.0 \ 10^{11}$ $L_\odot$ getting a total luminosity for the cD of 6.0 $10^{11}$ $L_\odot$ (we assume $M_B = 5.44$ for the Sun).

Adopting the lower value M/L $= 50$ for the halo and 21.9 for the core these translates respectively into a mass of $M(halo) = 5.0 \ 10^{12}$ $M_\odot$, $M(core) = 1.1 \ 10^{13}$ $M_\odot$ and a total mass for the cD of 1.6 $10^{13}$ $M_\odot$. This means that while the halo supplies about 17% of the total light of the cD it contributes by at least 1/3 to the total mass of the galaxy. This fraction increases to 2/3 if we had adopted instead a halo M/L$_B = 200$ (a total mass of 3.1 $10^{13}$ $M_\odot$ would thus result) as the population synthesis model would lead to prefer.

It is hard to say, at the present status, whether the halo formed by accreting intracluster matter in the gaseous phase as the cooling-flows theory would suggest or rather by stripping mass in stellar form via cannibalism processes involving other nearby galaxies. We simply note in this regard that our estimate of the present total mass of the cD halo would require the formation (cooling flows) from 30 to 120 galaxies or to cannibalize them within an Hubble time.

*Acknowledgements.* The project was partly funded by EEC Contract ERB-CHRX-CT92-0093.